\documentclass{article}
\usepackage{booktabs}
\usepackage{spconf,amsmath,graphicx}
\usepackage{url}

\title{Enhancing GAN-Based Vocoders with Contrastive Learning Under Data-limited Condition}
\name{Author(s) Name(s)\thanks{Thanks to XYZ agency for funding.}}
\address{Author Affiliation(s)}

\name{Haoming Guo\thanks{This paper is based on Haoming's thesis \cite{thesis} at University of California, Berkeley. }, Seth Z. Zhao, Jiachen Lian, Gopala Anumanchipalli, Gerald Friedland}
\address{University of California, Berkeley \\
  {\tt\small \{mike0221, 
sethzhao506,
jiachenlian,
gopala,
fractor
\}@berkeley.edu}}
\begin{document}
%
\maketitle

\begin{abstract}
Vocoder models have recently achieved substantial progress in generating authentic audio comparable to human quality while significantly reducing memory requirement and inference time. However, these data-hungry generative models require large-scale audio data for learning good representations. In this paper, we apply contrastive learning methods in training the vocoder to improve the perceptual quality of the vocoder without modifying its architecture or adding more data. We design an auxiliary task with mel-spectrogram contrastive learning to enhance the utterance-level quality of the vocoder model under data-limited conditions. We also extend the task to include waveforms to improve the multi-modality comprehension of the model and address the discriminator overfitting problem. We optimize the additional task simultaneously with GAN training objectives. Our results show that the tasks improve model performance substantially in data-limited settings. 
\end{abstract}

\section{Introduction}
Generative Adversarial Networks (GANs)~\cite{GAN} have been widely used in vocoders and have achieved the state-of-the-art in the domain ~\cite{melgan, hifigan, lee2023bigvgan}. However, training GAN vocoders still meets two challenges, data insufficiency and discriminator overfitting.

In the realm of single-speaker speech synthesis, the limited size of available datasets poses a significant challenge. To enhance the performance of vocoders operating under such constraints, we propose the use of unsupervised learning techniques to extract additional self-supervised signals for training. Self-supervised learning (SSL) methods have demonstrated efficacy in a diverse array of speech domains, including representation learning \cite{oord2018representation-cpc, HuBert, chen2022wavlm, AudioContrastive, mae_that_listens, Titouan2023}, synthesis \cite{ni2022utts-2, lian2022robust,lian2022c-dsvae, lian2022utts}, and multi-modality \cite{shi2022learning-avhubert, lian2023av-data2vec}. Drawing on the exceptional transfer learning capabilities of SSL, we seek to harness this power in the realm of Vocoder modeling, focusing specifically on the application of contrastive learning. Although contrastive learning has been explored in the context of speech recognition~\cite{oord2018representation-cpc}, we are unaware of any previous efforts to apply this approach to Vocoder modeling. In this work, our aim is to leverage contrastive learning as an auxiliary task to enhance the vocoding performance of GAN generators under data-limited conditions. 

The second challenge, discriminator overfitting, is also shown to be crucial, especially on small dataset \cite{Zhao2020DifferentiableAF, Tseng2021RegularizingGA, NEURIPS2020_8d30aa96}, and the convergence of GAN also critically depends on the quality of discriminators \cite{ProjectedGAN}. Contrastive learning on the discriminator has been proved to alleviate this problem in image generation \cite{jeong2021training}, and the method, in general, is also shown to increase model's performance and robustness on vision and language tasks \cite{xue2022investigating, Hendrycks2019UsingSL, Ghosh_2021_CVPR, SMD}. However, in speech synthesis, a naive approach of mel-spectrogram contrastive learning will only involve the generator, which encodes mel-spectrograms, but not the discriminator, which encodes the waveform. Therefore, we propose to extend the training to the discriminator by using a multi-modal contrastive task between mel-spectrograms and waveforms. 

Our contributions can be summarized as the following. We propose a contrastive learning task that explores mel-spectrogram and waveform interactions in utterance level, which improves the fidelity of vocoder on limited data. This self-supervised learning objective could be integrated seamlessly into standard GAN-based vocoder training pipeline. Extensive experiments and in-depth analysis demonstrate the effectiveness of our method's effectiveness in low-resource scenarios.

\section{Methods} \label{methods}

In this section, we first introduce the auxiliary contrastive task that we have designed for the GAN vocoder model. Subsequently, we explicate the details of how we modified the task to train both the generator and the discriminator of the vocoder model. Finally, we illustrate our proposed training framework, which synergizes the contrastive task with GAN objectives. It is worth noting that we have utilized the same model architecture as HiFi-GAN \cite{hifigan}. However, it is pertinent to mention that our method can be applied to other GAN frameworks for vocoders as well.

\subsection{Mel-spectrogram Contrastive Learning}\label{subsec:gen}

\begin{figure*}
\begin{minipage}{\textwidth}
    \centering
    \includegraphics[width=0.9\linewidth]{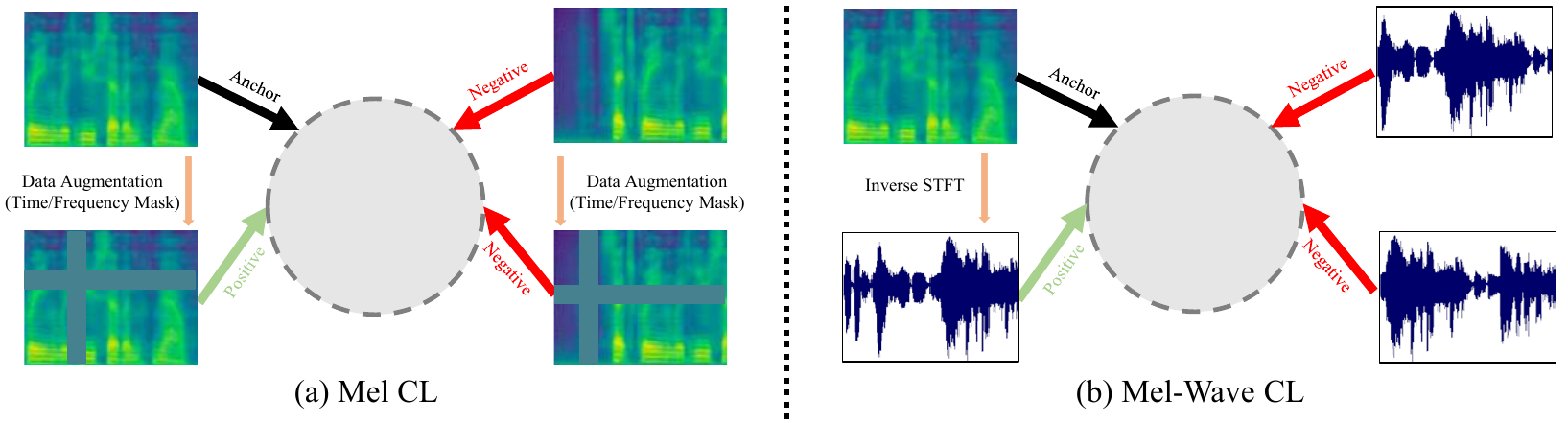}
    \vspace{-12pt}
    \caption{\textbf{Illustration of Contrastive Learning framework.} Our framework consists of two different formulations: (a) Mel-spectrogram Contrastive Learning (Mel CL); (b) Mel-Spectrogram \& Waveform Contrastive Learning (Mel-Wave CL). Notice that negative samples are randomly sampled inside each batch.}
    \label{fig:cl_frame}
\end{minipage}
\end{figure*}

In our GAN model, the generator takes a mel-spectrogram as input and outputs a raw waveform through a stack of convolutional layers. We use a learnable feed-forward layer to project the features of the convolutional layers onto a latent space $R^D$ , where elements of similar semantics are close to each other through contrastive learning. For each anchor in a batch of $N$ samples, we apply masking on randomly selected intervals in time and frequency to create a positive sample, while all other $(N - 1)$ input samples and $(N - 1)$ masked samples are used as negative samples. Together, the method results in $1$ positive pair and $2(N - 1)$ negative pairs in the batch. We then adapt the InfoNCE loss \cite{infonce} used in CLIP \cite{clip} for our loss function as follows:

\begin{equation}
\begin{split}
\mathcal{L}_{cl} & = - \frac{1}{N}\sum_{i=1}
^N\left(\log
\frac{\text{exp}(\tau {\textbf{v}_{i}} \cdot {\textbf{v}_{k})}}
{\sum_{j=1;i \neq j}^{2N} \text{exp}(\tau {\textbf{v}_{i}} \cdot {\textbf{v}_{j})})}
\right)
\end{split}
\end{equation}
where $\textbf{v}_{k} \in R^D$ is the masked sample from $\textbf{v}_{i} \in R^D$ and $\tau$ is a temperature parameter. This method is shown in Fig. \ref{fig:cl_frame}(a).

\subsection{Mel-spectrogram Waveform Contrastive Learning}\label{subsec:disc}

In addition to training solely the generator, we propose a novel task that involves contrastive spectrogram-waveform matching. This task serves to train both the generator and the discriminators, promoting rich semantic representation and preventing overfitting of the discriminators to the real or fake classification. The method is illustrated in Fig. \ref{fig:cl_frame}(b). For a batch of pairs of mel-spectrograms and waveforms, we assign the labels of the true pairs to be positive and those of the other pairs to be negative, resulting in $N$ positive pairs and $N(N-1)$ negative pairs in a batch of $N$ samples. We use the backbone of the generator to encode the mel-spectrogram and the backbone of the discriminator to encode the waveform. Similar to the method in section \ref{subsec:gen}, we use two separate feed-forward layers to project each encoded feature to the same latent dimension $R^D$. Then, we perform the modified loss function
\begin{equation}
\begin{split}
\mathcal{L}_{cl} & = - \frac{1}{N}\sum_{i=1}
^N\left(\log
\frac{\text{exp}(\tau {\textbf{v}_{i}} \cdot {\textbf{w}_{i})}}
{\sum_{j=1;i \neq j}^{N} \text{exp}(\tau {\textbf{v}_{i}} \cdot {\textbf{w}_{j})})}
\right)
\end{split}
\end{equation}
where $\textbf{w}_{i} \in R^D$ is the latent embedding of the waveform corresponding to the $i$th mel-spectrogram,  $\textbf{v}_{i} \in R^D$ is the latent embedding of the $i$th mel-spectrogram, and $\tau$ is a temperature parameter. HiFi-GAN contains multiple discriminators, so we calculate a contrastive loss between the mel-spectrogram embedding and each of the waveform embeddings and sum them up. For simplicity, we refer them as one discriminator in this paper unless otherwise mentioned. 


\subsection{Multi-tasking Framework}
\begin{figure}[t]
    \centering
    \includegraphics[width=1.0\linewidth]{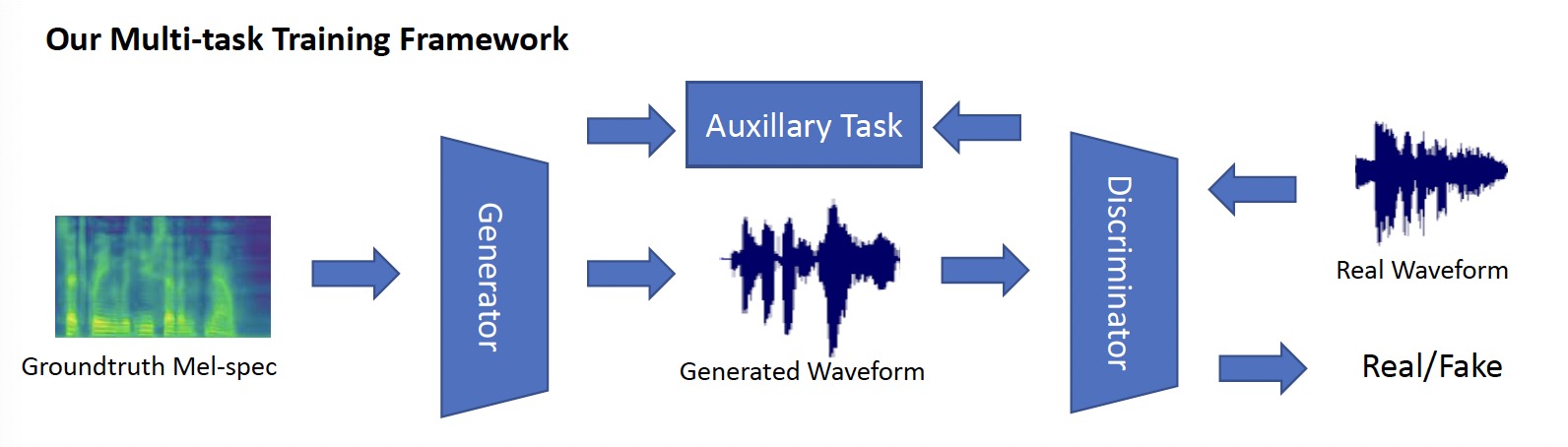}
    \vspace{-24pt}
    \caption{\textbf{Illustration of our multi-tasking frameworks.} To incorporate the auxiliary contrastive learning task, we propose a  multi-tasking framework, in which we set the contrastive task as additional learning objectives along with the original GAN optimization objectives. This framework applies to both contrastive learning methods described in section \ref{subsec:gen} and \ref{subsec:disc}.}
    \label{fig:framework}
\end{figure}

To integrate contrastive learning with GAN tasks, we adopt a multi-tasking framework that makes auxiliary tasks a joint optimization objective with original learning goals \cite{aux_1}. As illustrated in Fig. \ref{fig:framework}, we create additional heads for the training generator and discriminator with auxiliary tasks. The total loss for training the vocoder model thus becomes: 
\begin{equation}
\begin{split}
\mathcal{L}_{G} = \mathcal{L}_{adv} + \lambda_{fm}\mathcal{L}_{fm} + \lambda_{mel}\mathcal{L}_{mel} + 
\lambda_{cl} \mathcal{L}_{cl}
\end{split}
\end{equation}
\begin{equation}
\begin{split}
\mathcal{L}_{D} = \mathcal{L}_{adv} + \mathcal{I}_{disc}\lambda_{cl} \mathcal{L}_{cl}
\end{split}
\end{equation}
where $\mathcal{L}_{G}$ is the total loss for the generator and $\mathcal{L}_{D}$ is the total loss for the discriminator. $\mathcal{L}_{adv}$ is the adversarial loss,  $\mathcal{L}_{fm}$ is the feature matching loss, and $\mathcal{L}_{mel}$ is the mel-spectrogram reconstruction loss in the original HiFi-GAN training pipeline. $\mathcal{L}_{mel}$ can be either of the contrastive loss described in section \ref{subsec:gen} or \ref{subsec:disc}, and $\mathcal{I}_{disc}$ is an indicator of whether the latter is used. Each loss is weighted with a $\lambda$ coefficient which can be set as hyperparameters. We use a $\lambda_{fm}$ of 2, $\lambda_{mel}$ of 45 from the HiFi-GAN setting \cite{hifigan} and a $\lambda_{cl}$ of 1.

\section{Experiments}
\subsection{Experimental Setting}

In this section, we describe the details of our experimental settings including the dataset, model choice, hyperparameters and evaluation metrics.

\subsubsection{Dataset}
In order to have a fair comparison with other vocoder models, we train the model on the LJSpeech dataset \cite{ljspeech17} which is also used in other vocoder works like HiFi-GAN \cite{hifigan}. LJSpeech is a public single-speaker dataset with 13100 short English audio clips whose durations span from 1 second to 10 seconds. We use the default data split with 12950 training samples and 150 validation samples. We use the same preprocessing configurations with HiFi-GAN, including 80 bands of mel-spectrograms as input and FFT size of 1024, window size of 1024, and hop size of 256 for conversion from waveform to mel-spectrograms.\cite{hifigan}

\subsubsection{Implementation details}
For experimental comparison on audio quality, we choose the most powerful HiFi-GAN V1 and the most lightweight HiFi-GAN V3 as the baseline methods, and we use the same model architecture as the backbone to apply the contrastive tasks described in section \ref{subsec:gen} and \ref{subsec:disc}. Under the multi-tasking framework, we train HiFi-GAN along with the contrastive learning methods with a batch size of 16, an AdamW optimizer, and a learning rate of 0.0002. For the following experiments on the full dataset, all models are trained for 400k steps on one Nvidia TITAN RTX GPU. The experiments on 20\% of the dataset train for 300k steps on the same device, and those on 4\% and 0.8\% of the dataset train for 200k and 40k steps, respectively. The model inference time on GPU is about 70ms for V1 models and 32ms for V3 models. 

\begin{table}[h]
  \centering
  \resizebox{0.7\linewidth}{!}{
  \begin{tabular}{l|cc|c}
    \toprule[1.2pt]
    Model & MAE & MCD & MOS (CI)\\
    \midrule
    Ground Truth & - & - & 4.32 ($\pm0.05$)\\
    \midrule
    HiFi-GAN V1 & \textbf{0.111} & \textbf{4.203} & \textbf{4.21} ($\pm0.05$) \\
    + Mel CL & 0.114 & 4.289 & 4.18 ($\pm0.06$)\\
    + Mel-Wave CL & 0.113 & 4.228 & 4.20 ($\pm0.05$)\\
    \midrule
    HiFi-GAN V3 & \textbf{0.203} & 7.786 & 4.10 ($\pm0.05$) \\
    + Mel CL & 0.204 & 7.766 & \textbf{4.13 ($\pm0.07$)}\\
    + Mel-Wave CL & \textbf{0.203} & \textbf{7.723} & 4.09 ($\pm0.06$)\\
  \bottomrule[1.2pt]
\end{tabular}}
\caption{Objective and subjective evaluation results for models with mel-spectrogram contrastive loss (Mel CL) and mel-spectrogram contrastive loss (Mel-Wave CL). Models are trained on the full training set. CI is 95\% confidence interval of the MOS score.}
\label{tab:results_full}
\end{table}
\vspace{-12pt}

\begin{table*}[h]
  \centering
  \resizebox{0.7\linewidth}{!}{
  \begin{tabular}{l|cc|c}
    \toprule[1.2pt]
    Model & MAE $\downarrow$ & MCD $\downarrow$ & MOS $\uparrow$ (CI)\\
    \midrule
    Ground Truth & - & - & 4.32 ($\pm0.05$)\\
    \midrule
    HiFi-GAN V1 (20\% data) & \textbf{0.113} $(\uparrow0.002)$ & 4.352 $(\uparrow0.149)$ & 4.13 $(\downarrow0.08)$ ($\pm0.06$) \\
    + Mel CL (20\% data) & 0.116 $(\uparrow0.002)$ & 4.430 $(\uparrow0.139)$ & 4.11 $(\downarrow0.07)$ ($\pm0.07$) \\
    + Mel-Wave CL (20\% data) & \textbf{0.113} $(\uparrow0.000)$ & \textbf{4.295} $(\uparrow0.067)$ & \textbf{4.16} $(\downarrow0.04)$ ($\pm0.06$)\\
    \midrule
    Hifi-GAN V3 (20\% data) & 0.212 $(\uparrow0.009)$ & 8.157 $(\uparrow0.371)$ & 3.88 $(\downarrow0.22)$ ($\pm0.06$)\\
    + Mel CL (20\% data) & \textbf{0.207} $(\uparrow0.003)$ & \textbf{7.960} $(\uparrow0.206)$ & 3.95 $(\downarrow0.18)$ ($\pm0.06$)\\
    + Mel-Wave CL (20\% data) & \textbf{0.207} $(\uparrow0.004)$ & 7.974 $(\uparrow0.251)$ & \textbf{4.04} $(\downarrow0.05)$ ($\pm0.07$)\\
    \midrule
    \midrule
    HiFi-GAN V1 (4\% data) & 0.137 $(\uparrow0.026)$ & 5.372 $(\uparrow1.169)$ & 3.80 $(\downarrow0.41)$ ($\pm0.05$) \\
    + Mel-Wave CL (4\% data) & 0.135 $(\uparrow0.022)$ & 5.201 $(\uparrow0.973)$ & 3.86 $(\downarrow0.34)$ ($\pm0.06$) \\
    \midrule
    HiFi-GAN V1 (0.8\% data) & 0.205 $(\uparrow0.094)$ & 7.912 $(\uparrow3.709)$ & 3.48 $(\downarrow0.73)$ ($\pm0.12$) \\
    + Mel-Wave CL (0.8\% data)  & 0.188 $(\uparrow0.075)$ & 7.125 $(\uparrow2.897)$ & 3.63 $(\downarrow0.57)$ ($\pm0.09$)\\
  \bottomrule[1.2pt]
\end{tabular}}
\caption{Objective and subjective evaluation results for models trained with different percentages of the training set. The number in parenthesis indicates the difference from the results when trained on the full dataset. Notice that our method suffers from less fidelity degradations compared to the baseline method under various data-limited scenarios.}
\label{tab:results_partial}
\end{table*}

\subsubsection{Evaluation metrics}
To objectively evaluate our models compared to the baseline, we measure the mean average error (MAE) and mel-cepstral distortion (MCD) \cite{Kubichek1993MelcepstralDM} on mel-spectrograms. On both metrics, lower scores indicate closer alignment with the ground truth. We also include a 5-scale mean opinion score (MOS) on audio quality as a subjective evaluation performed on 50 samples excluded from the training set.

\subsection{Results} 

We present the results of models trained on full data with the multi-tasking framework in Table \ref{tab:results_full}. Below, we refer Mel CL as the mel-spectrogram contrastive learning in section \ref{subsec:gen}, and
Mel-Wave CL as the mel-spectrogram waveform contrastive learning in section \ref{subsec:disc}. For V1 models, the baseline performs slightly better than the proposed methods by margins of 0.02 on MAE, 0.025 on MCD, and 0.01 on MOS. For V3 models, on the objective tests, we observe that the model trained with mel-spectrogram contrastive loss has comparable performance with the baseline, while the one trained with mel-spectrogram waveform contrastive loss achieves the highest scores on both metrics. The results show that our proposed methods have at least comparable performance to the baseline HiFi-GAN when training on the full dataset. On the subjective tests, the V3 model with Mel CL achieves the highest MOS score, 0.03 above the V3 baseline. The model with Mel-Wave CL has a similar MOS score with the baseline on the full dataset. Overall, when trained on the full dataset, the proposed methods have limited gains on top of the baseline.

To investigate how each model performs under data limitation, we train the three models on 20\% of the dataset and evaluate them with the same validation set. We present the results in Table \ref{tab:results_partial}. With less data, the baseline HiFi-GAN V3 suffers a significant performance degradation across all metrics, including 0.371 on MCD and 0.22 on MOS. Meanwhile, the V3 model trained with Mel CL experiences an increase of 0.194 on MCD and a drop of 0.18 on MOS. The V3 model trained with Mel-Wave CL has an increase of 0.251 on MCD and a drop of only 0.05 on MOS. It suggests Mel-Wave CL is most resistant to data insufficiency. The two proposed methods have comparable scores on the objective evaluation, but the model with Mel-Wave CL obtains a significantly higher score on the subjective test, 0.16 higher than the V3 baseline. The findings align with our hypothesized alleviation of discriminator overfitting by Mel-Wave CL, which is a more severe problem on the small training dataset. Both of the proposed methods perform substantially better than the baseline by 0.07 and 0.16 respectively. 

A similar trend exists in the HiFi-GAN V1 experiments, where Mel-Wave CL achieves the best scores and the least performance drop on all metrics. One slightly surprising finding is that the larger model V1 often experiences a smaller performance drop compared to the smaller model V3 when trained on 20\% data. Typically, a larger model is expected to be more prone to overfitting when trained on less data, which should lead to a larger performance drop. In this specific case, however, HiFi-GAN V1 has a larger generator but the same discriminator as HiFi-GAN V3 \cite{hifigan}, which is our suspected reason for the finding. Overall, the results show the benefits of additional supervision signals from contrastive learning in data-limited situations and the superior performance of Mel-Wave CL on a small dataset.

Since Mel-Wave CL demonstrates significant improvement over the baselines, we run more extreme cases of training on only 4\% and 0.8\% of the training set (513 and 104 training samples, respectively) to further validate its usefulness. The results are shown in table \ref{tab:results_partial}. Mel-Wave CL still outperforms the baseline V1 by significant margins on all metrics, which shows its consistency in improving the model in data-limited situations.


One possible source of improvement of our framework is data augmentation, but we argue that using contrastive learning is better than direct data augmentation. With direct augmentation, the mel-spectrogram loss in HiFi-GAN training would require the model to map masked and unmasked spectrograms to the same waveform, causing conflicts in generator’s upsampling. Contrastive learning leverages data augmentation without messing up the mel-spectrogram loss. Our experiment on HiFi-GAN V1+SpecAugment yields MAE of 0.122, MCD of 4.515 and MOS of 4.03 when training on full dataset, significantly worse than the baseline and both our contrastive methods. The performance degradation is also supported by other studies \cite{lee2023bigvgan}.

\section{Conclusion} \label{conclusion}
This paper describes our proposed contrastive learning framework to improve GAN vocoders. Our results show the legacy of using contrastive learning as an auxiliary task that facilitates vocoder training without adding more data or modifying model architecture. We demonstrate that the proposed framework significantly outperforms the baseline when training on limited data by extracting additional supervision signals and reducing discriminator overfitting. 

For future work, we plan to repeat the experiments on different datasets to test our method's generalizability. In particular, we want to test its extension to multi-speaker datasets, another domain where data insufficiency is critical.  
{\small
\bibliographystyle{IEEEbib}
\bibliography{refs}

\begin{thebibliography}{10}

\bibitem{thesis}
Haoming Guo, Gerald Friedland, and Gopala~Krishna Anumanchipalli,
\newblock ``Enhancing gan-based vocoders with contrastive learning,''
\newblock M.S. thesis, EECS Department, University of California, Berkeley, May
  2023.

\bibitem{GAN}
Ian~J. Goodfellow, Jean Pouget-Abadie, Mehdi Mirza, Bing Xu, David
  Warde-Farley, Sherjil Ozair, Aaron Courville, and Yoshua Bengio,
\newblock ``Generative adversarial networks,'' 2014.

\bibitem{melgan}
Kundan Kumar, Rithesh Kumar, Thibault de~Boissiere, Lucas Gestin, Wei~Zhen
  Teoh, Jose Sotelo, Alexandre de~Br\'{e}bisson, Yoshua Bengio, and Aaron~C
  Courville,
\newblock ``Melgan: Generative adversarial networks for conditional waveform
  synthesis,''
\newblock in {\em Advances in Neural Information Processing Systems},
  H.~Wallach, H.~Larochelle, A.~Beygelzimer, F.~d\textquotesingle
  Alch\'{e}-Buc, E.~Fox, and R.~Garnett, Eds. 2019, vol.~32, Curran Associates,
  Inc.

\bibitem{hifigan}
Jungil Kong, Jaehyeon Kim, and Jaekyoung Bae,
\newblock ``Hifi-gan: Generative adversarial networks for efficient and high
  fidelity speech synthesis,''
\newblock {\em ArXiv}, vol. abs/2010.05646, 2020.

\bibitem{lee2023bigvgan}
Sang gil Lee, Wei Ping, Boris Ginsburg, Bryan Catanzaro, and Sungroh Yoon,
\newblock ``Big{VGAN}: A universal neural vocoder with large-scale training,''
\newblock in {\em The Eleventh International Conference on Learning
  Representations}, 2023.

\bibitem{oord2018representation-cpc}
Aaron van~den Oord, Yazhe Li, and Oriol Vinyals,
\newblock ``Representation learning with contrastive predictive coding,''
\newblock {\em arXiv preprint arXiv:1807.03748}, 2018.

\bibitem{HuBert}
Wei-Ning Hsu, Benjamin Bolte, Yao-Hung~Hubert Tsai, Kushal Lakhotia, Ruslan
  Salakhutdinov, and Abdelrahman Mohamed,
\newblock ``Hubert: Self-supervised speech representation learning by masked
  prediction of hidden units,'' 2021.

\bibitem{chen2022wavlm}
Sanyuan Chen, Chengyi Wang, Zhengyang Chen, Yu~Wu, Shujie Liu, Zhuo Chen, Jinyu
  Li, Naoyuki Kanda, Takuya Yoshioka, Xiong Xiao, et~al.,
\newblock ``Wavlm: Large-scale self-supervised pre-training for full stack
  speech processing,''
\newblock {\em IEEE Journal of Selected Topics in Signal Processing}, vol. 16,
  no. 6, pp. 1505--1518, 2022.

\bibitem{AudioContrastive}
Aaqib Saeed, David Grangier, and Neil Zeghidour,
\newblock ``Contrastive learning of general-purpose audio representations,''
  2020.

\bibitem{mae_that_listens}
Po-Yao Huang, Hu~Xu, Juncheng Li, Alexei Baevski, Michael Auli, Wojciech
  Galuba, Florian Metze, and Christoph Feichtenhofer,
\newblock ``Masked autoencoders that listen,'' 2022.

\bibitem{Titouan2023}
Titouan Parcollet, Shucong Zhang, Rogier van Dalen, Alberto Gil C.~P. Ramos,
  and Sourav Bhattacharya,
\newblock ``{On the (In)Efficiency of Acoustic Feature Extractors for
  Self-Supervised Speech Representation Learning},''
\newblock in {\em {Interspeech 2023}}, Dublin, France, Aug. 2023.

\bibitem{ni2022utts-2}
Junrui Ni, Liming Wang, Heting Gao, Kaizhi Qian, Yang Zhang, Shiyu Chang, and
  Mark Hasegawa-Johnson,
\newblock ``Unsupervised text-to-speech synthesis by unsupervised automatic
  speech recognition,''
\newblock {\em arXiv preprint arXiv:2203.15796}, 2022.

\bibitem{lian2022robust}
Jiachen Lian, Chunlei Zhang, and Dong Yu,
\newblock ``Robust disentangled variational speech representation learning for
  zero-shot voice conversion,''
\newblock in {\em ICASSP 2022-2022 IEEE International Conference on Acoustics,
  Speech and Signal Processing (ICASSP)}. IEEE, 2022, pp. 6572--6576.

\bibitem{lian2022c-dsvae}
Jiachen Lian, Chunlei Zhang, Gopala~Krishna Anumanchipalli, and Dong Yu,
\newblock ``Towards improved zero-shot voice conversion with conditional
  dsvae,''
\newblock {\em arXiv preprint arXiv:2205.05227}, 2022.

\bibitem{lian2022utts}
Jiachen Lian, Chunlei Zhang, Gopala~Krishna Anumanchipalli, and Dong Yu,
\newblock ``Utts: Unsupervised tts with conditional disentangled sequential
  variational auto-encoder,''
\newblock {\em arXiv preprint arXiv:2206.02512}, 2022.

\bibitem{shi2022learning-avhubert}
Bowen Shi, Wei-Ning Hsu, Kushal Lakhotia, and Abdelrahman Mohamed,
\newblock ``Learning audio-visual speech representation by masked multimodal
  cluster prediction,''
\newblock {\em arXiv preprint arXiv:2201.02184}, 2022.

\bibitem{lian2023av-data2vec}
Jiachen Lian, Alexei Baevski, Wei-Ning Hsu, and Michael Auli,
\newblock ``Av-data2vec: Self-supervised learning of audio-visual speech
  representations with contextualized target representations,''
\newblock {\em arXiv preprint arXiv:2302.06419}, 2023.

\bibitem{Zhao2020DifferentiableAF}
Shengyu Zhao, Zhijian Liu, Ji~Lin, Jun-Yan Zhu, and Song Han,
\newblock ``Differentiable augmentation for data-efficient gan training,''
\newblock {\em ArXiv}, vol. abs/2006.10738, 2020.

\bibitem{Tseng2021RegularizingGA}
Hung-Yu Tseng, Lu~Jiang, Ce~Liu, Ming-Hsuan Yang, and Weilong Yang,
\newblock ``Regularizing generative adversarial networks under limited data,''
\newblock {\em 2021 IEEE/CVF Conference on Computer Vision and Pattern
  Recognition (CVPR)}, pp. 7917--7927, 2021.

\bibitem{NEURIPS2020_8d30aa96}
Tero Karras, Miika Aittala, Janne Hellsten, Samuli Laine, Jaakko Lehtinen, and
  Timo Aila,
\newblock ``Training generative adversarial networks with limited data,''
\newblock in {\em Advances in Neural Information Processing Systems},
  H.~Larochelle, M.~Ranzato, R.~Hadsell, M.F. Balcan, and H.~Lin, Eds. 2020,
  vol.~33, pp. 12104--12114, Curran Associates, Inc.

\bibitem{ProjectedGAN}
Axel Sauer, Kashyap Chitta, Jens M{\"{u}}ller, and Andreas Geiger,
\newblock ``Projected gans converge faster,''
\newblock {\em CoRR}, vol. abs/2111.01007, 2021.

\bibitem{jeong2021training}
Jongheon Jeong and Jinwoo Shin,
\newblock ``Training {\{}gan{\}}s with stronger augmentations via contrastive
  discriminator,''
\newblock in {\em International Conference on Learning Representations}, 2021.

\bibitem{xue2022investigating}
Yihao Xue, Kyle Whitecross, and Baharan Mirzasoleiman,
\newblock ``Investigating why contrastive learning benefits robustness against
  label noise,''
\newblock in {\em First Workshop on Pre-training: Perspectives, Pitfalls, and
  Paths Forward at ICML 2022}, 2022.

\bibitem{Hendrycks2019UsingSL}
Dan Hendrycks, Mantas Mazeika, Saurav Kadavath, and Dawn~Xiaodong Song,
\newblock ``Using self-supervised learning can improve model robustness and
  uncertainty,''
\newblock in {\em Neural Information Processing Systems}, 2019.

\bibitem{Ghosh_2021_CVPR}
Aritra Ghosh and Andrew Lan,
\newblock ``Contrastive learning improves model robustness under label noise,''
\newblock in {\em Proceedings of the IEEE/CVF Conference on Computer Vision and
  Pattern Recognition (CVPR) Workshops}, June 2021, pp. 2703--2708.

\bibitem{SMD}
Kehan Wang, Seth~Z. Zhao, David Chan, Avideh Zakhor, and John Canny,
\newblock ``Multimodal semantic mismatch detection in social media posts,''
\newblock in {\em Proceedings of IEEE 24th International Workshop on Multimedia
  Signal Processing (MMSP)}, 2022.

\bibitem{infonce}
Aaron van~den Oord, Yazhe Li, and Oriol Vinyals,
\newblock ``Representation learning with contrastive predictive coding,'' 2018.

\bibitem{clip}
Alec Radford, Jong~Wook Kim, Chris Hallacy, Aditya Ramesh, Gabriel Goh,
  Sandhini Agarwal, Girish Sastry, Amanda Askell, Pamela Mishkin, Jack Clark,
  Gretchen Krueger, and Ilya Sutskever,
\newblock ``Learning transferable visual models from natural language
  supervision,'' 2021.

\bibitem{aux_1}
Fengda Zhu, Yi~Zhu, Xiaojun Chang, and Xiaodan Liang,
\newblock ``Vision-language navigation with self-supervised auxiliary reasoning
  tasks,'' 2019.

\bibitem{ljspeech17}
Keith Ito and Linda Johnson,
\newblock ``The lj speech dataset,''
  \url{https://keithito.com/LJ-Speech-Dataset/}, 2017.

\bibitem{Kubichek1993MelcepstralDM}
Robert~F. Kubichek,
\newblock ``Mel-cepstral distance measure for objective speech quality
  assessment,''
\newblock {\em Proceedings of IEEE Pacific Rim Conference on Communications
  Computers and Signal Processing}, vol. 1, pp. 125--128 vol.1, 1993.

\end{thebibliography}
}
\end{document}